\documentclass[a4paper]{panl}
\usepackage{cite}
\usepackage{wrapfig}
\usepackage{graphicx}
\usepackage{amssymb}
\usepackage{amsfonts}
\usepackage{amsmath}
\usepackage{longtable}
\usepackage{rotating}
\usepackage{lscape}
\usepackage{epsfig}
\usepackage{color}

\originalTeX
\begin{document}
\issuearea{Physics of Elementary Particles and Atomic Nuclei. Theory}

\title{$E_6$ inspired composite Higgs model and baryon asymmetry generation}
\maketitle
\authors{R.\,Nevzorov${}^{a,}$\footnote{E-mail: nevzorov@itep.ru},
A.~W.\,Thomas${}^{b}$}
\from{$^{a}$\,NRC Kurchatov Institute --- ITEP, Moscow, 117218, Russia}
\vspace{-3mm}
\from{$^{b}$\,ARC Centre of Excellence for Particle Physics at the Terascale and CSSM,
Department of Physics, The University of Adelaide, Adelaide SA 5005, Australia}
%
%
%
%

\begin{abstract}{
The breakdown of $SU(6)$ global symmetry down to its $SU(5)$ subgroup near the scale $f\gtrsim 10\,\mbox{TeV}$
in the strongly interacting sector within the $E_6$ inspired composite Higgs model (E$_6$CHM) gives rise to
a set of pseudo--Nambu--Goldstone bosons (pNGBs) that involves one Standard Model (SM) singlet scalar,
a SM--like Higgs doublet and an $SU(3)_C$ triplet of scalar fields, $T$. We argue that the baryon number violation
in the E$_6$CHM can induce the observed matter–antimatter asymmetry if CP is violated. The coloured triplet
of scalar fields with mass in the few TeV range plays a key role in this process and may lead to a distinct new physics
signal that can be detected at the LHC in the near future.
}
\end{abstract}
\vspace*{6pt}


\section*{Introduction}

The observed baryon asymmetry in the Universe stimulates the exploration of different extensions of the Standard Model (SM).
This asymmetry can be created dynamically within the scenarios satisfying Sakharov conditions \cite{Sakharov:1967dj}.
A number of such new physics scenarios were proposed including GUT baryogenesis \cite{gut-baryogen-1,gut-baryogen-2},
baryogenesis via leptogenesis \cite{Fukugita:1986hr}, the Affleck-Dine mechanism \cite{Affleck-Dine-1},
electroweak baryogenesis \cite{ew-baryogen}, etc\,. Here we consider the generation of the baryon asymmetry within the $E_6$
inspired composite Higgs model (E$_6$CHM) \cite{Nevzorov:2015sha,Nevzorov:2016fxp,Nevzorov:2016jba,Nevzorov:2017rtf,Nevzorov:2018fug}.

The idea of a composite Higgs boson, that was proposed in the 70's \cite{Terazawa:1976xx-1} and
80's \cite{composite-higgs-1}, implies that there exists a new, strongly coupled sector.
This sector induces the electroweak (EW) scale dynamically, in analogy with the origin of the QCD scale.
The minimal composite Higgs model (MCHM) \cite{Agashe:2004rs} involves two sectors (for a review, see Ref.~\cite{Bellazzini:2014yua}).
The first sector contains weakly-coupled elementary particles, including all SM gauge bosons and SM fermions.
The second sector gives rise to a set of bound states that involves the Higgs doublet and massive fields with
the quantum numbers of all SM particles which are associated with the composite partners of the quarks, leptons and gauge bosons.

At low energies the elementary and composite states with the same quantum numbers mix, so that the SM fermions (bosons)
are superpositions of the corresponding elementary fermion (boson) states and their composite fermion (boson) partners.
In this partial compositeness framework \cite{Contino:2006nn, Kaplan:1991dc} the SM fields couple to the Higgs boson
and other composite states with a strength determined by the compositeness fraction of each SM field.
In this case the observed mass hierarchy in the quark and lepton sectors can be accommodated if the fractions of compositeness
of the first and second generation fermions are quite small. This also results in the suppression of off-diagonal flavor
transitions, as well as modifications of the $W$ and $Z$ couplings associated with these light fermion fields \cite{Contino:2006nn}.
At the same time the right-handed top quark $t^c$ should have sizeable fraction of compositeness because the top quark is rather heavy.

In the MCHM the Higgs doublet emerges as a set of pseudo--Nambu--Goldstone bosons (pNGBs) from the spontaneous breakdown of
an approximate global $\mbox{SO(5)}$ symmetry of the strongly coupled sector. Near the scale $f$ the $\mbox{SO(5)}$ symmetry is broken
down to $\mbox{SO(4)} \cong SU(2)_W\times SU(2)_R$ so that the $\mbox{SU(2)}_W\times U(1)_Y$ subgroup of the SM gauge group remains intact.
The global custodial symmetry $\mbox{SU(2)}_{cust} \subset \mbox{SU(2)}_W\times \mbox{SU(2)}_R$ protects the Peskin-Takeuchi $\hat{T}$ parameter
against the contributions induced by the composite states. Nevertheless experimental limits on the parameter $|\hat{S}|\lesssim 0.002$
imply that the set of spin-1 resonances, which, in particular, includes the composite partners of the SM gauge bosons, should be heavier than
$2.5\,\mbox{TeV}$.

On the other hand, even more stringent bounds come from the observed suppression of the non--diagonal flavor transitions.
Although partial compositeness considerably reduces the contributions of composite states to the corresponding processes, it was shown that
in general the adequate suppression of the non--diagonal flavor transitions can be achieved only if $f$ is larger than $10\,\mbox{TeV}$.
However in the models with additional flavour symmetries $\mbox{FS}=U(2)^3=U(2)_{q}\times U(2)_u \times U(2)_d$,
under which the first two generations of elementary quark states transform as doublets and the third generation as singlets,
this bound on the scale $f$ can be substantially alleviated. For low values of the scale $f$ the appropriate suppression of
the Majorana masses of the left-handed neutrinos and the baryon number violating operators may be attained provided global $U(1)_B$ and
$U(1)_L$ symmetries, which guarantee the conservation of the baryon and lepton numbers respectively, are imposed.

In the E$_6$CHM the strongly interacting sector possesses an approximate $SU(6)\times U(1)_L$ symmetry \cite{Nevzorov:2015sha}.
The global $SU(6)$ symmetry is expected to be broken down to $SU(5)$, which contains the SM gauge group, near the scale, $f\gtrsim 10\,\mbox{TeV}$.
Such breakdown results in a set of pNGBs that, in particular, includes the SM--like Higgs doublet.
The observed baryon asymmetry can be generated in this case if CP and $U(1)_B$ are violated.
In the next section the E$_6$CHM is briefly reviewed. The process of the baryon asymmetry generation is considered in the second
last section. Last section is reserved for our conclusions.

\section*{$E_6$ inspired composite Higgs model}

The $E_6$ inspired composite Higgs model (E$_6$CHM) can arise after the breakdown of gauge symmetry
within the $N=1$ supersymmetric (SUSY) orbifold Grand Unified Theories (GUTs) in six dimensions which
are based on the $E_6\times G_0$ gauge group \cite{Nevzorov:2015sha}. Near some high energy scale, $M_X$,
the $E_6\times G_0$ symmetry can be broken down to the $SU(3)_C\times SU(2)_W\times U(1)_Y \times G$ subgroup,
where $SU(3)_C\times SU(2)_W\times U(1)_Y$ is the SM gauge group. Multiplets from the strongly interacting
sector are charged under both the $E_6$ and $G_0$ ($G$) gauge symmetries. The elementary states from
the weakly--coupled sector participate in the $E_6$ interactions only. In the SUSY orbifold GUTs under consideration
different multiplets of the elementary quarks and leptons can stem from different fundamental $27$-dimensional
representations of $E_6$. All other components of the corresponding $27$--plets should gain masses of the order of $M_X$.
It is expected that in this case SUSY is broken somewhat below the GUT scale $M_X$
(Different phenomenological aspects of the $E_6$ inspired models with low-scale SUSY breaking were explored in
\cite{Hall:2010ix,Athron:2010zz,Nevzorov:2012hs,Nevzorov:2013tta,Nevzorov:2013ixa,Athron:2014pua,Athron:2015vxg,Athron:2016gor}.).

All fields from the strongly interacting sector in these orbifold GUTs are localised on the brane, where
the $E_6$ symmetry is broken down to the $SU(6)$ that contains $SU(3)_C\times SU(2)_W\times U(1)_Y$ subgroup.
As a consequence at high energies the Lagrangian of the composite sector respects $SU(6)$ symmetry.
Although the SM gauge interactions violate this global symmetry, $SU(6)$ can remain an
approximate global symmetry of the strongly coupled sector, even at low energies. This happens when
the gauge couplings of the composite sector are substantially larger than the SM gauge couplings.
Hereafter we assume that around the scale $f\gtrsim 10\,\mbox{TeV}$ the $SU(6)$ global symmetry
is broken down to its $\mbox{SU(5)}$ subgroup, so that the SM gauge group is preserved.
The corresponding breakdown of $\mbox{SU(6)}$ gives rise to a set of pNGB states that includes the SM--like Higgs doublet.

The global $U(1)_L$ global symmetry suppresses the operators in the composite sector of the E$_6$CHM
that induce too large Majorana masses of the left--handed neutrino. To ensure that the left--handed neutrinos
gain small Majorana masses this symmetry has to be broken down to
\begin{equation}
Z^L_{2}=(-1)^{L} \, ,
\label{1}
\end{equation}
in the weakly--coupled sector, where $L$ is a lepton number. Also it is assumed that the low energy effective
Lagrangian of the E$_6$CHM is invariant with respect to an approximate $Z^B_2$ symmetry,
which is a discrete subgroup of $U(1)_{B}$, i.e.
\begin{equation}
Z^B_{2}=(-1)^{3B} \, ,
\label{3}
\end{equation}
where $B$ is the baryon number. The $Z^L_2$ and $Z^B_2$ discrete symmetries forbid operators that
lead to rapid proton decay. All other baryon and lepton number violating operators are sufficiently
strongly suppressed by the large value of the scale $f$, as well as small mixing between elementary
states and their composite partners \cite{Nevzorov:2017rtf,Nevzorov:2018fug}.

In order to embed the E$_6$CHM into orbifold GUTs based on the $E_6\times G_0$ gauge group,
the SM gauge couplings extrapolated to high energies using the renormalisation group equations (RGEs)
have to converge to some common value near the scale $M_X$. Within the E$_6$CHM an approximate gauge
coupling unification can be achieved if the right--handed top quark $t^c$ is composite.
In this case the weakly--coupled sector includes \cite{Nevzorov:2015sha}
\begin{equation}
(q_i,\,d^c_i,\,\ell_i,\,e^c_i) + u^c_{\alpha} + \bar{q}+\bar{d^c}+\bar{\ell}+\bar{e^c}\,,
\label{4}
\end{equation}
where $\alpha=1,2$ and $i=1,2,3$. In Eq.~(\ref{4}) $q_i$ and $\ell_i$ are doublets of the left-handed quarks
left-handed lepton, $e_i^c$ are the right--handed charged leptons, $u^c_i$ and $d^c_j$ are the right-handed
up- and down-type quarks, whereas the extra exotic states $\bar{q},\,\bar{d^c},\,\bar{\ell}$ and $\bar{e^c}$,
have opposite $SU(3)_C\times SU(2)_W\times U(1)_Y$ quantum numbers to the left-handed quark doublets, right-handed
down-type quarks, left-handed lepton doublets and right-handed charged leptons, respectively.
The set of fermion states (\ref{4}) is chosen so that the weakly--coupled sector includes all SM fermions
except $t^c$ and anomaly cancellation takes place.

It is rather easy to find the value of $\alpha_3(M_Z)$ that results in the exact
gauge coupling unification in the E$_6$CHM for $\alpha(M_Z)=1/127.9$ and $\sin^2\theta_W=0.231$.
In the one--loop approximation one obtains
\begin{equation}
\dfrac{1}{\alpha_3(M_Z)}=\dfrac{1}{b_1-b_2}\biggl[\dfrac{b_1-b_3}{\alpha_2(M_Z)}-
\dfrac{b_2-b_3}{\alpha_1(M_Z)}\biggr]\,,
\label{41}
\end{equation}
where $b_i$ are one--loop beta functions, with the indices $1,\,2,\,3$ corresponding
to the $U(1)_Y$, $SU(2)_W$ and $SU(3)_C$ interactions. Because all composite states come in complete
$SU(5)$ multiplets, the strongly interacting sector does not contribute to the differential running determined
by $(b_i-b_j)$. Then the elementary particle spectrum (\ref{4}) leads to the exact gauge coupling unification, which
takes place near $M_X\sim 10^{16}\, \mbox{GeV}$, if $\alpha_3(M_Z)\simeq 0.109$\,. This result indicates that
for $\alpha_3(M_Z)\simeq 0.118$ the SM gauge couplings may be rather close to each other at high energies.

The E$_6$CHM implies that the strongly coupled sector gives rise to the composite fermions that
form ${\bf 10} + {\bf \overline{5}}$ multiplets of $SU(5)$. These fermions get combined with
$\bar{q},\,\bar{d^c},\,\bar{\ell}$ and $\bar{e^c}$, resulting in a set of vector--like states.
The only exceptions are the components of the $10$--plet associated with $t^c$, which survive down to
the EW scale. In the simplest scenario the composite ${\bf 10} + {\bf \overline{5}}$ multiplets of $SU(5)$
stem from one ${\bf{15}}$--plet and two ${\bf \overline{6}}$--plets (${\bf \overline{6}}_1$ and ${\bf \overline{6}}_2$)
of $SU(6)$ that decompose under $SU(3)_C\times SU(2)_W\times U(1)_Y$ as follows:
\begin{equation}
\begin{array}{ll}
\begin{array}{rcl}
{\bf 15} &\to& Q = \left(3,\,2,\,\dfrac{1}{6}\right)\,,\\[0mm]
&& t^c = \left(3^{*},\,1,\,-\dfrac{2}{3}\right)\,,\\[0mm]
&& E^c = \Biggl(1,\,1,\,1\Biggr)\,,\\[0mm]
&& D = \left(3,\,1,\,-\dfrac{1}{3} \right)\,,\\[0mm]
&& \overline{L}=\left(1,\,2,\,\dfrac{1}{2}\right)\,;
\end{array}
\qquad
\noindent
\begin{array}{rcl}
{\bf \overline{6}}_{\alpha} &\to & D^c_{\alpha} = \left(\bar{3},\,1,\,\dfrac{1}{3} \right)\,,\\[0mm]
& & L_{\alpha} = \left(1,\,2,\,-\dfrac{1}{2} \right)\,,\\[0mm]
& & N_{\alpha} = \Biggl(1,\,1,\,0 \Biggr)\,,
\end{array}
\end{array}
\label{5}
\end{equation}
where $\alpha=1,2$. The first and second quantities in brackets are the $SU(3)_C$ and $SU(2)_W$ representations,
whereas the third ones are the $U(1)_Y$ charges. The large mass of the top quark can be induced only in the case when
$t^c$ is $Z^B_2$-odd state. Therefore all components of the ${\bf{15}}$--plet should be odd under the $Z^B_2$ symmetry.
At the same time we assume the components of ${\bf \overline{6}}_1$ and ${\bf \overline{6}}_2$ are $Z^B_2$-even and
$Z^B_2$-odd respectively.

After the $SU(6)$ symmetry breaking $\bar{q},\,\bar{d^c},\,\bar{\ell}$, $\bar{e^c}$ as well as all composite
fermions (\ref{5}) except $t^c$, $N_1$ and $N_2$ form vector--like states with masses which are larger than $f$.
The breakdown of the $SU(6)$ symmetry should also induce Majorana masses for the SM singlet states $N_1$ and $N_2$.
The mixing between $N_1$ and $N_2$ has to be suppressed because of the approximate $Z^B_2$ symmetry.
In the next section we discuss the generation of the baryon asymmetry, assuming that $N_1$ is considerably
lighter than other exotic fermion states and has a mass which is somewhat smaller than $f$.

\section*{Generation of matter–antimatter asymmetry}

The breakdown of the $SU(6)$ to its $SU(5)$ subgroup gives rise to gives rise to eleven pNGB states.
These pNGB states consist of one $SU(2)_W$ doublet associated with the SM--like Higgs doublet $H$,
one $SU(3)_C$ triplet of scalar fields $T$ and one real SM singlet scalar. The masses of the pNGB states
tend to be considerably lower than $f$. Therefore pNGBs are the lightest composite resonances in the E$_6$CHM
and may lead to distinct collider signatures \cite{Nevzorov:2016fxp, Nevzorov:2016jba}.

All pNGBs have to be $Z^B_2$-even states because Higgs doublet is $Z^B_2$-even.
As a result the $SU(3)_C$ scalar triplet can decay into up and down antiquarks.
Since the fractions of compositeness of the first and second generation quarks are rather small
the decay mode $T\to\bar{t}\bar{b}$ should be the dominant one. At low energies $E\ll f$ all baryon
number violating operators are strongly suppressed and $T$ manifests itself in the interactions
with other SM particles as a diquark. At the LHC, the $\mbox{SU(3)}_C$ scalar triplet can be pair
produced, leading to some enhancement of the cross section for the process $pp\to T\bar{T}\to t\bar{t}b\bar{b}$.

In the E$_6$CHM the Lagrangian, that describes the interactions of $N_1$ and $N_2$ with $T$
and down-type quarks, is given by
\begin{equation}
\mathcal{L}_{N}= \sum_{i=1}^3 \Biggl( g^{*}_{i1} T d^c_i N_1 + g^{*}_{i2} T d^c_i N_2 + h.c.\Biggr) \, .
\label{6}
\end{equation}
In the exact $Z^B_2$ symmetry limit, $g_{i1}=0$. The approximate $Z^B_2$ symmetry
implies that $|g_{i1}| \ll |g_{i2}|$. In the model under consideration
the matter–antimatter asymmetry can be induced via the out--of equilibrium decays
$N_{1}\to T+\bar{d}_i$ and $N_{1}\to T^{*}+ d_i$, provided $N_1$ is the lightest composite
fermion in the spectrum \cite{Nevzorov:2017rtf,Nevzorov:2018fug}.

The process of matter–antimatter asymmetry generation is controlled by the three
CP (decay) asymmetries
\begin{equation}
\varepsilon_{1,\,k}=\dfrac{\Gamma_{N_1 d_{k}}-\Gamma_{N_1 \bar{d}_{k}}}
{\sum_{m} \left(\Gamma_{N_1 d_{m}}+\Gamma_{N_1 \bar{d}_{m}}\right)} \, ,
\label{7}
\end{equation}
where $\Gamma_{N_1 d_{k}}$ and $\Gamma_{N_1 \bar{d}_{k}}$ are partial decay widths
of $N_1\to d_k + T^{*}$ and $N_1\to \overline{d}_k + T$ with $k,m=1,2,3$.
At the tree level these decay asymmetries vanish. The non--zero contributions
to the CP asymmetries (\ref{7}) come from the interference between the tree--level
amplitudes of the $N_1$ decays and the one--loop corrections to them.
Because $T$ couples mainly to the third generation fermions so that $|g_{31}|\gg |g_{21}|,\, |g_{11}|$
and $|g_{32}|\gg |g_{22}|,\, |g_{12}|$, the decay asymmetries $\varepsilon_{1,\,2}$ and
$\varepsilon_{1,\,1}$ are much smaller than $\varepsilon_{1,\,3}$ and
can be ignored in the leading approximation. Assuming that
$N_1$ is substantially lighter than other composite fermion states including $N_2$,
the computation of one--loop diagrams gives \cite{Nevzorov:2017rtf,Nevzorov:2018fug}
\begin{equation}
\varepsilon_{1,\,3}\simeq -\dfrac{1}{(4\pi)}\dfrac{|g_{32}|^2}{\sqrt{x}}\sin 2\Delta\varphi\,,\qquad\qquad
\Delta\varphi=\varphi_{32}-\varphi_{31}\,,
\label{11}
\end{equation}
where $x=(M_2/M_1)^2$, $M_1$ and $M_2$ are the Majorana masses of $N_1$ and $N_2$, $g_{31}=|g_{31}| e^{i\varphi_{31}}$
and $g_{32}=|g_{32}| e^{i\varphi_{32}}$. Here it is assumed that the mass of the $SU(3)_C$ scalar triplet
$m_{T}\ll M_1$ and therefore can be neglected. The CP asymmetry $\varepsilon_{1,\,3}$ reaches its absolute maximum
value for $\Delta\varphi=\pm \pi/4$ and vanishes if CP invariance is preserved, i.e. all Yukawa couplings are real.

To compute the total baryon asymmetries induced by the decays of $N_1$, the system of Boltzmann equations
that describe the evolution of baryon number densities needs to be solved. Since the corresponding solution
should be similar to the solutions of the Boltzmann equations for thermal leptogenesis, the generated
matter–antimatter asymmetry can be estimated using an approximate formula \cite{Nevzorov:2017rtf,Nevzorov:2018fug}
\begin{equation}
Y_{\Delta B}\sim 10^{-3}\biggl(\varepsilon_{1,\,3} \eta_3\biggr)\,,
\label{10}
\end{equation}
where $\eta_3$ is an efficiency factor, that varies from 0 to 1, and
$Y_{\Delta B}$ is the baryon asymmetry relative to the entropy density $s$, i.e.
\begin{equation}
Y_{\Delta B}=\dfrac{n_B-n_{\bar{B}}}{s}\biggl|_0=(8.75\pm 0.23)\times 10^{-11}\,.
\label{9}
\end{equation}
A thermal population of $N_1$ decaying completely out of equilibrium without
washout effects would lead to $\eta_{3}=1$. Washout processes reduce the induced
matter–antimatter asymmetry by the factors $\eta_3$. Here we ignore sphaleron
interactions that should partially convert baryon asymmetry into lepton asymmetry.

In the E$_6$CHM $\eta_3$ can be of the order of unity.
Indeed, in the strong washout scenario this efficiency factor may be estimated as follows
\begin{equation}
\begin{array}{c}
\eta_3 \simeq H(T=M_1)/\Gamma_{3}\,,\\[3mm]
\Gamma_3 = \Gamma_{N_1 d_{3}}+\Gamma_{N_1 \bar{d}_{3}}=\dfrac{3 |g_{31}|^2}{16 \pi}\,M_1\,,
\qquad\qquad
H=1.66 g_{*}^{1/2}\dfrac{T^2}{M_{Pl}}\,,
\end{array}
\label{12}
\end{equation}
where $H$ is the Hubble expansion rate and $g_{*}=n_b+\dfrac{7}{8}\,n_f$ is the number of relativistic
degrees of freedom in the thermal bath. Within the SM $g_{*}=106.75$, while in the E$_6$CHM
$g_{*}=113.75$ for $T\lesssim f$. From Eqs.~(\ref{12}) it follows that $\eta_3$ increases
with diminishing of $|g_{31}|$ and for $M_1\simeq 10\,\mbox{TeV}$ it becomes close to unity
when $|g_{31}|\sim 10^{-6}$.

If $\eta_3 \sim 1$, the induced baryon asymmetry is determined by the CP asymmetry
$\varepsilon_{1,\,3}$ which is set by $|g_{32}|$ and the combination of the CP--violating phases $\Delta \varphi$.
For $(M_2/M_1)=10$ a phenomenologically acceptable value of the baryon density, corresponding
to $\varepsilon_{1,\,3}\lesssim 10^{-7}$ can be generated even when $|g_{32}|\lesssim 0.01$
On the other hand because the Yukawa coupling of $N_2$ to $SU(3)_C$ scalar triplet and $b$-quark is not suppressed
by the $Z_2^B$ symmetry, $|g_{32}|$ is expected to be relatively large, i.e. $|g_{32}| \gtrsim 0.1$.
For so large values of $|g_{32}|$ and $\Delta\varphi \simeq \pi/4$ the phenomenologically
acceptable baryon density can be obtained only if $\eta_3\lesssim 10^{-3}$. When $\eta_3 \sim 1$
and $|g_{32}|\gtrsim 0.1$ the observed baryon density can be generated for rather small values
of the CP--violating phase $\Delta\varphi\lesssim 0.01$. This demonstrates that the appropriate matter–antimatter
asymmetry can be obtained within the E$_6$CHM even if CP is approximately preserved.

\section*{Conclusions}

The gauge symmetry breaking within $N=1$ SUSY orbifold GUTs which are based on the $E_6\times G_0$ gauge group
may lead to the $E_6$ inspired composite Higgs model (E$_6$CHM) with approximate $SU(6)$ symmetry in the
strongly interacting sector. In the E$_6$CHM the right-handed top quark is a composite state. To ensure
its possible embedding into a suitable GUTs this model contains a set of exotic fermions that, in particular,
involves two SM singlet Majorana states $N_1$ and $N_2$. Within the E$_6$CHM the operators that give rise to
rapid proton decay are suppressed by a $Z^L_{2}$ discrete symmetry. Near the scale $f\gtrsim 10\,\mbox{TeV}$
the $SU(6)$ symmetry is supposed to be broken down to its $SU(5)$ subgroup, which incorporates
the SM gauge group, leading to a set of pNGB states that, in particular, compose the SM--like Higgs doublet and
$SU(3)_C$ triplet of scalar fields, $T$. In general the pNGBs form the lightest composite resonances with masses
which are substantially lower than $f$, whereas all exotic fermions and other composite states acquire masses
which are somewhat larger than $f$. Assuming that $N_1$ is the lightest exotic fermion, with a mass around
$10\,\mbox{TeV}$, we argued that in the E$_6$CHM with baryon number violation the observed baryon asymmetry
can be generated via the out--of equilibrium decays $N_{1}\to T+\bar{b}$ and $N_{1}\to T^{*}+ b$, provided CP
is violated. Moreover, if the absolute value of the Yukawa coupling of $N_2$ to $T$ and $b$--quark is larger than $0.1$,
a phenomenologically acceptable baryon density may be obtained, even when all CP--violating phases are rather small
($\lesssim 0.01$). The $SU(3)_C$ scalar triplet $T$, with mass in the few TeV range, can be pair produced at the LHC and
predominantly decays into $T\to \bar{t}+\bar{b}$, resulting in the enhancement of the cross section
of $pp\to t\bar{t}b\bar{b}$.


\begin{thebibliography}{99}
\bibitem{Sakharov:1967dj}
{\it Sakharov A.D.} Violation of CP invariance, C asymmetry, and baryon asymmetry of the universe //
JETP Lett. 1967. V. 5. P. 24.
\bibitem{gut-baryogen-1}
{\it Ignatiev A.Yu., Krasnikov N.V., Kuzmin V.A., Tavkhelidze A.N.} Universal CP noninvariant superweak
interaction and baryon asymmetry of the universe // Phys. Lett. B 1978. V. 76. P. 436.
\bibitem{gut-baryogen-2}
{\it Yoshimura M.} Unified gauge theories and the baryon number of the Universe // Phys. Rev. Lett. 1978. V. 41. P. 281.
\bibitem{Fukugita:1986hr}
{\it Fukugita M., Yanagida T.} Baryogenesis Without Grand Unification // Phys. Lett. B 1986. V. 174. P. 45.
\bibitem{Affleck-Dine-1}
{\it Affleck I., Dine M.} A new mechanism for baryogenesis // Nucl. Phys. B 1985. V. 249. P. 361.
\bibitem{ew-baryogen}
{\it Riotto A., Trodden M.} Recent progress in baryogenesis // Ann. Rev. Nucl. Part. Sci. 1999. V. 49. P. 35.
\bibitem{Nevzorov:2015sha}
{\it Nevzorov R., Thomas A.W.} $E_6$ inspired composite Higgs model // Phys. Rev. D 2015. V. 92. P. 075007.
\bibitem{Nevzorov:2016fxp}
{\it Nevzorov R., Thomas A.W.} LHC signatures of neutral pseudo-Goldstone boson in the $E_6$CHM //
J. Phys. G 2017. V. 44. No. 7. P. 075003.
\bibitem{Nevzorov:2016jba}
{\it Nevzorov R., Thomas A.W.} $E_6$ inspired composite Higgs model and 750 GeV diphoton excess //
EPJ Web Conf. 2016. V. 125. P. 02021.
\bibitem{Nevzorov:2017rtf}
{\it Nevzorov R., Thomas A.W.} Baryon asymmetry generation in the E$_6$CHM // Phys. Lett. B
2017. V. 774. P. 123.
\bibitem{Nevzorov:2018fug}
{\it Nevzorov R., Thomas A.W.} Generation of baryon asymmetry in the E$_6$CHM // EPJ Web Conf. 2018. V. 191. P. 02004.
\bibitem{Terazawa:1976xx-1}
{\it Terazawa H., Akama K., Chikashige Y.} Unified Model of the Nambu-Jona-Lasinio Type for All Elementary Particle Forces //
Phys. Rev. D 1977. V. 15. P. 480.
\bibitem{composite-higgs-1}
{\it Dimopoulos S., Preskill J.} Massless Composites With Massive Constituents // Nucl. Phys. B 1982. V. 199. P. 206.
\bibitem{Agashe:2004rs}
{\it Agashe K., Contino R., Pomarol A.} The Minimal composite Higgs model // Nucl. Phys. B 2005. V. 719. P. 165.
\bibitem{Bellazzini:2014yua}
{\it Bellazzini B., Csáki C., Serra J.} Composite Higgses // Eur. Phys. J. C 2014. V. 74. No. 5. P. 2766.
\bibitem{Contino:2006nn}
{\it Contino R., Kramer T., Son M., Sundrum R.} Warped/composite phenomenology simplified // JHEP 2007. V. 0705. P. 074.
\bibitem{Kaplan:1991dc}
{\it Kaplan D. B.} Flavor at SSC energies: A New mechanism for dynamically generated fermion masses // Nucl. Phys. B 1991. V. 365. P. 259.
\bibitem{Hall:2010ix}
{\it Hall J. P., King S. F., Nevzorov R., Pakvasa S., Sher M.} Novel Higgs Decays and Dark Matter in the E$_6$SSM // Phys. Rev. D 2011. V. 83. P. 075013.
\bibitem{Athron:2010zz}
{\it Athron P., Hall J. P., Howl R., King S. F., Miller D. J., Moretti S., Nevzorov R.} Aspects of the exceptional supersymmetric standard model // Nucl. Phys. Proc. Suppl. 2010. V. 200-202. P. 120.
\bibitem{Nevzorov:2012hs}
{\it Nevzorov R.} $E_6$ inspired supersymmetric models with exact custodial symmetry // Phys. Rev. D 2013. V. 87. No. 1. P. 015029.
\bibitem{Nevzorov:2013tta}
{\it Nevzorov R., Pakvasa S.} Exotic Higgs decays in the $E_6$ inspired SUSY models // Phys. Lett. B 2014. V. 728. P. 210.
\bibitem{Nevzorov:2013ixa}
{\it Nevzorov R.} Quasifixed point scenarios and the Higgs mass in the E6 inspired supersymmetric models // Phys. Rev. D 2014. V. 89. No. 5. P. 055010.
\bibitem{Athron:2014pua}
{\it Athron P., Mühlleitner M., Nevzorov R., Williams A. G.} Non-Standard Higgs Decays in U(1) Extensions of the MSSM // JHEP 2015. V. 1501. P. 153.
\bibitem{Athron:2015vxg}
{\it Athron P., Harries D., Nevzorov R., Williams A. G.} $E_6$ Inspired SUSY benchmarks, dark matter relic density and a 125 GeV Higgs // Phys. Lett. B
2016. V. 760. P. 19.
\bibitem{Athron:2016gor}
{\it Athron P., Harries D., Nevzorov R., Williams A. G.} Dark matter in a constrained E$_{6}$ inspired SUSY model // JHEP 2016. V. 1612. P. 128.
\end{thebibliography}
\end{document}